%% file: revised_2.tex
\documentclass[superscriptaddress, reprint, aip, jcp, floatfix]{revtex4-2}

\usepackage{xcolor}

\usepackage{graphicx}% Include figure files
\usepackage{dcolumn}% Align table columns on decimal point
\usepackage{bm}% bold math
\usepackage[utf8]{inputenc}
\usepackage[T1]{fontenc}
\usepackage{mathptmx}
\usepackage{etoolbox}
\usepackage[normalem]{ulem}
\makeatletter
\def\@email#1#2{%
 \endgroup
 \patchcmd{\titleblock@produce}
  {\frontmatter@RRAPformat}
  {\frontmatter@RRAPformat{\produce@RRAP{*#1\href{mailto:#2}{#2}}}\frontmatter@RRAPformat}
  {}{}
}%
\makeatother
\begin{document}

\preprint{AIP/123-QED}

\title{Entanglement as a topological marker in harmonically confined attractive Fermi-Hubbard chains}

\author{M. Sanino, I. M. Carvalho, and V. V. Fran\c{c}a}

\affiliation{S\~ao Paulo State University (UNESP), Institute of Chemistry, 14800-090, Araraquara, S\~{a}o Paulo, Brazil}

\begin{abstract}

We investigate the single-site von Neumann entropy along a harmonically confined superfluid chain, as described by the one-dimensional fermionic Hubbard model with strongly attractive interactions. We find that by increasing the confinement (or equivalently the particle filling) the system undergoes a quantum phase transition from a superfluid (SF) to a non-trivial topological insulator (TI) phase, which is characterized by an insulating bulk surrounded by highly entangled superfluid edges. These highly entangled states are found to be robust against perturbations and topologically protected by the particle-hole symmetry, which is locally preserved. We also find a semi-quantitative agreement between entanglement and superconducting order parameter profiles, confirming then that entanglement can be used as a topological marker and an order parameter in these systems. The charge gap not only confirms the SF-TI transition, but also shows that this transition is mediated by a metallic intermediate regime within the bulk. Using the gap we could depict a phase diagram for which the non-trivial topological insulator can be found in such harmonically confined attractive systems. 
\end{abstract}

\maketitle

%\begin{quotation}

%\end{quotation}
\vspace{-0.2cm}
\section{Introduction}
\vspace{-0.2cm}
Topological states have received interest across various physical systems due to their unique properties, such as robustness against local perturbations and disorder \cite{PhysRevB.109.094202, WANG2024112111, liao2024visualizing, doi:10.1126/sciadv.aax2007, doi:10.1126/sciadv.aat3187,CONTINENTINO2017A1, olaniyan2024switchable}. These properties not only deepen our understanding of quantum materials, but can also be used in practical applications, including fault-tolerant quantum computing and novel electronic or photonic devices \cite{Lima2023,lee2024fault, dai2024programmable}.

Most of the studies exploring fundamental differences between topological and conventional insulators are in {\bf k} space. For noninteracting systems, topological states are in many cases well described via topological band theory~\cite{ZAHIDHASAN2013143}, based on the assumption of perfect translational symmetry \cite{haldane1988model, kane2006new}, which allows for a well-defined global Chern number. 

Nevertheless, a local description of topological order has been proposed in coordinate space~\cite{bianco2011mapping} for independent spinless electrons in 2D systems. This {\it local} Chern number not only shows that topological insulators present a peculiar distribution of electrons, but also allows one to explore topological properties even in the absence of a well-defined {\bf k}-space, as in inhomogeneous systems. A
similar approach has been proposed to the statistical Boltzmann entropy for the three-dimensional optical lattices~\cite{paiva2011fermions}.

In interacting systems, correlations and many-body effects can significantly alter topological properties, making it challenging to define topological invariants, particularly in systems without translational symmetry, such as 1D optical lattices~\cite{PhysRevB.108.195151,bianco2011mapping, bandres2016topological, benalcazar2017electric}. On the other hand, state-of-the-art experiments with ultracold atoms in optical lattices~\cite{Schafer2020,doi:10.1126/science.aal3837} offer a promising platform for exploring topological quantum phenomena. Notably, they enable the realization of the fermionic Hubbard model~\cite{lewenstein2012ultracold}, where interactions and harmonic confinement play a crucial role. While the effects of harmonic trapping have been extensively examined in bosonic and fermionic systems ~\cite{PhysRevA.97.023605,PhysRevA.100.053611,PhysRevLett.93.200402}, its implications for topological states remain unexplored.

In this context, entanglement --- which has been proved to be a powerful tool for probing quantum phase transitions \cite{korepin2004universality, zozulya2007bipartite, laflorencie2016quantum, Bloch2008,doi:10.1126/science.aaz6801, Pauletti2024,PhysRevLett.93.250404,RevModPhys.80.517, pauletti2024linear, canella2019superfluid, canella2020entanglement, arisa2020linear, canella2022effects} ---  emerges as a potential indicator of topological order. Since entanglement captures fundamental nonlocal correlations, typically driven by charge and spin fluctuations, its analysis can reveal unconventional arrangements of electrons. For example, the entanglement spectrum has been used to fingerprint topological order \cite{li2008entanglement, oliveira2014entanglement, brzezinska2018entanglement}. In Kitaev materials spin-orbit entanglement actually leads to topological phases \cite{TREBST20221}. Entanglement entropy\cite{RevModPhys.80.517} has also been used to determine topological order in spin-liquid phase in a Bose$-$Hubbard model on the Kagome lattice \cite{isakov2011topological}.  

Here we explore the single-site {entanglement} entropy along a harmonically confined chain described by the 1D Hubbard model in the strongly attractive regime. This {\it entanglement profile} --- which to our knowledge has never been used to characterize quantum phase transitions --- has the potential to directly probe the non-trivial topological structure within the system. We find that by increasing the harmonic strength (or equivalently by increasing the particle filling), the system undergoes a transition from a superfluid to a non-trivial topological insulator, characterized by an insulating bulk surrounded by highly entangled superfluid edges. We show that these highly entangled superfluid edges are robust against perturbations and are topologically protected by the particle-hole symmetry, which is locally preserved. The charge gap not only confirms our interpretation but also reveal a precursor metallic phase, allowing us to depict a phase diagram for superfluid, metal and non-trivial topological insulator phases as a function of density and confinement.  The entanglement profile is also found to be semi-quantitatively equivalent to the superconducting order parameter profile, confirming then that entanglement can be used as a topological marker and an order parameter in these systems. 
\vspace{-0.7cm}
\section{Model and Methods}
\vspace{-0.2cm}
We consider 1D fermionic Hubbard~\cite{giamarchi2003quantum} chains under harmonic confinement, described by the Hamiltonian
\vspace{-0.3cm}
\begin{eqnarray}
\hat{H} = &-&t\sum_{i=1,\sigma}^{L-1}(\hat{c}^{\dagger}_{i,\sigma}\hat{c}_{i+1, \sigma} + {\rm H.c.})\nonumber+U\sum_{i=1}^{L}\hat{n}_{i,\uparrow}\hat{n}_{i,\downarrow}\\
&+&k\sum_{i=1,\sigma}^{L}(i-(L+1)/2)^{2}\hat{n}_{i,\sigma},  
  \label{hamiltonian}
\end{eqnarray}
where $t$ is the nearest-neighboor hopping term,  $U<0$ is the attractive onsite interaction and $k$ is the curvature of the parabolic potential. Here $\hat{c}^{(\dagger)}_{i,\sigma}$ annihilates (creates) an electron with spin $\sigma=\uparrow,\downarrow$ at site $i$, $\hat{n}_{i,\sigma} = \hat{c}^{\dagger}_{i,\sigma}\hat{c}_{i,\sigma}$ is the number operator and $L$ is the chain size with open boundary conditions. Throughout this work, we set $t=1$ as the unit of energy. We here focus on strongly attractive interactions, $U=-10$, while the weakly attractive regime, including the BCS  limit~\cite{PhysRevB.55.575,RevModPhys.85.1633}, will be explored elsewhere.

The ground-state properties are obtained via Density Matrix Renormalization Group (DMRG) methods~\cite{SCHOLLWOCK_MPS,dmrg1} at a fixed filling $n=N/L$ and balanced spin populations $N_\uparrow=N_\downarrow=N/2$, where $N=\sum_{i}(\left\langle \hat{n}_{i,\uparrow} \right\rangle + \left\langle \hat{n}_{i,\downarrow}\right\rangle)=N_\uparrow + N_\downarrow$ is the total number of particles. The DMRG algorithm was implemented using the ITensor Library~\cite{itensor}, based on the matrix product states (MPS) ansatz~\cite{SCHOLLWOCK_MPS}. The accuracy of the MPS representation is controlled by the bond dimension, which was set to a maximum value of 2048. We initialize the DMRG calculations by configuring an initial state with doubly occupied sites, extending from the middle of the chain towards the boundaries. This approach allows faster convergence, as the target ground state exhibits a similar fermion distribution within the attractive regime. Our calculations were performed until the ground-state energy has converged to at least $10^{-7}$.

The density profile $\{\langle\hat{n}_i\rangle\}$ and the occupation probabilities' profiles $\{w_{i2},w_{i\uparrow}, w_{i\downarrow}, w_{i0} \}$ are directly obtained from the DMRG calculations. Here $w_{i2}$ is the double occupation at site $i$, $w_{i\sigma}$ the probability of single occupation with $\sigma$-spin  (for zero magnetization $w_{i\uparrow}=w_{i\downarrow}$), while $w_{i0}=1-w_{i\uparrow}-w_{i\downarrow}-w_{i2}$ is the probability of zero occupation. 

Entanglement between any bipartite system in a pure state is well quantified via the von Neumann entropy \cite{RevModPhys.81.865,RevModPhys.80.517}. There are several possible bipartitions that one could consider in our system, as for example entanglement between particles \cite{PhysRevB.92.075423,PhysRevB.105.115145} or entanglement between spatial regions of the chain \cite{pauletti2024linear,pauletti2024linear,PhysRevB.106.195405,zanardi,PhysRevLett.95.196406}. We explore  the single-site entanglement $S_i$, i.e. the entanglement between site $i$ and the remaining sites in the ground state, is calculated via the von Neumann entropy,
\begin{eqnarray}
    S_i&=&\frac{-1}{\log_2 d}{\rm Tr}[\rho_i\log_2 \rho_i] \nonumber\\
    &=&\frac{-1}{\log_2 d}\left[2w_{i\sigma}\log_2 w_{i\sigma}+w_{i2}\log_2 w_{i2}+w_{i0}\log_2 w_{i0}\right],
    \label{Si_entanglement}
\end{eqnarray}
where $d=4$ represents the single-site Hilbert space dimension and $\rho_i={\rm Tr}_{L-i}[\rho]$ is the reduced density matrix of $i$th site, obtained by tracing out the degrees of freedom of the remaining $L-i$ sites from the total density matrix $\rho$. As the terms $w_i$'s are the occupation probabilities at the site $i$, satisfying the relation $w_{i\uparrow} + w_{i\downarrow}+w_{i0}+w_{i2}=1$, entanglement is null when the site is in a well-defined state (any $w_i=1$). In the opposite scenario, whenever the site state is maximally mixed (all $w_i=1/4$) entanglement is maximum, $S_i=1$.

In general one has considered the average single-site entanglement, $\bar S=1/L\sum_i S_i$, for detecting quantum phase transitions~\cite{Coe_2011,PhysRevLett.100.070403,PhysRevA.83.042311,PhysRevA.81.052321,francca2017entanglement, canella2021mott,  PhysRevA.66.032110,doi:10.1126/science.aau0818,PhysRevB.73.085113,francca2017entanglement}. {It then quantifies the average nonlocal correlations between the electronic states of a single site and the remaining states of the system.} We here explore instead the {\it entanglement profile} $\{S_i\}$, {i. e. the spatial distribution of the single-site entanglement along the chain}, which to our knowledge has never been used to characterize quantum phase transitions. Notice that the difference between the average entanglement and the entanglement profile proposed here is, in a certain sense, analogous to the distinction between global and local Chern numbers. While the average (global) assigns a single value to the entire system, the profile (local) can capture the underlying organization of electrons. 

We also analyse the superconducting order parameter \cite{PhysRevLett.122.077002,PhysRevX.10.031016,Huang2022}, defined as
\begin{equation}
    \Delta_{pair}=\langle \hat{c}_{i,\downarrow}\hat{c}_{i,\uparrow}\hat{c}^\dagger_{i+1,\downarrow}\hat{c}^\dagger_{i+1,\uparrow}+{\rm H.c}.\rangle,
    \label{delta_pair}
\end{equation}
which accounts for the pair-pair correlations, and the charge gap,
\begin{eqnarray}
    \Delta_c&=&E_{\rm GS}(N-2)+E_{\rm GS}(N+2)-2E_{\rm GS}(N),
\end{eqnarray}
where $E_{\rm GS}(N)$ is the ground-state energy of a system with $N$ fermions with null magnetization.

\section{Results and discussion}\label{sec_results}

\begin{figure}[tb]
%\begin{centering}
\centering
\includegraphics[scale=0.48]{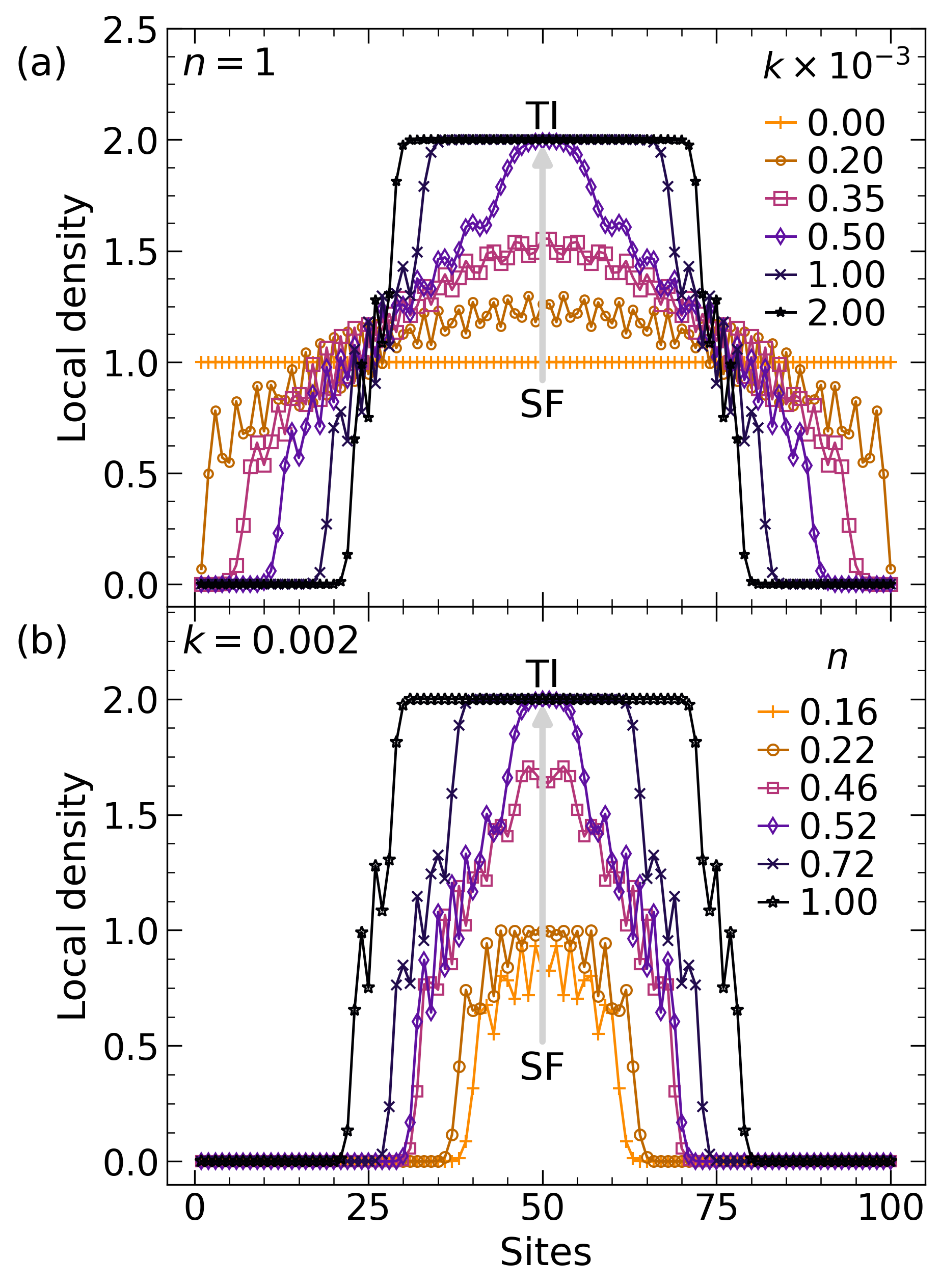}
%\par\end{centering}
\caption{(a) Effect of the harmonic curvature $k$ on the density profile at fixed $n=1$. As $k$ increases the effective chain at the core reduces, leading the superfluid (SF) bulk to an insulator phase (topological insulator, TI). As increasing $k$ is equivalent to increase the effective density, this superfluid-insulator transition at the bulk can also be driven by the average density $n$ at a fixed $k$, as shown in (b) for $k=0.002$. In all cases $L=100$ and $U=-10$. } \label{fig1}
\end{figure}

\begin{figure*}[tb]
%\begin{centering}
\centering
\includegraphics[scale=0.3]{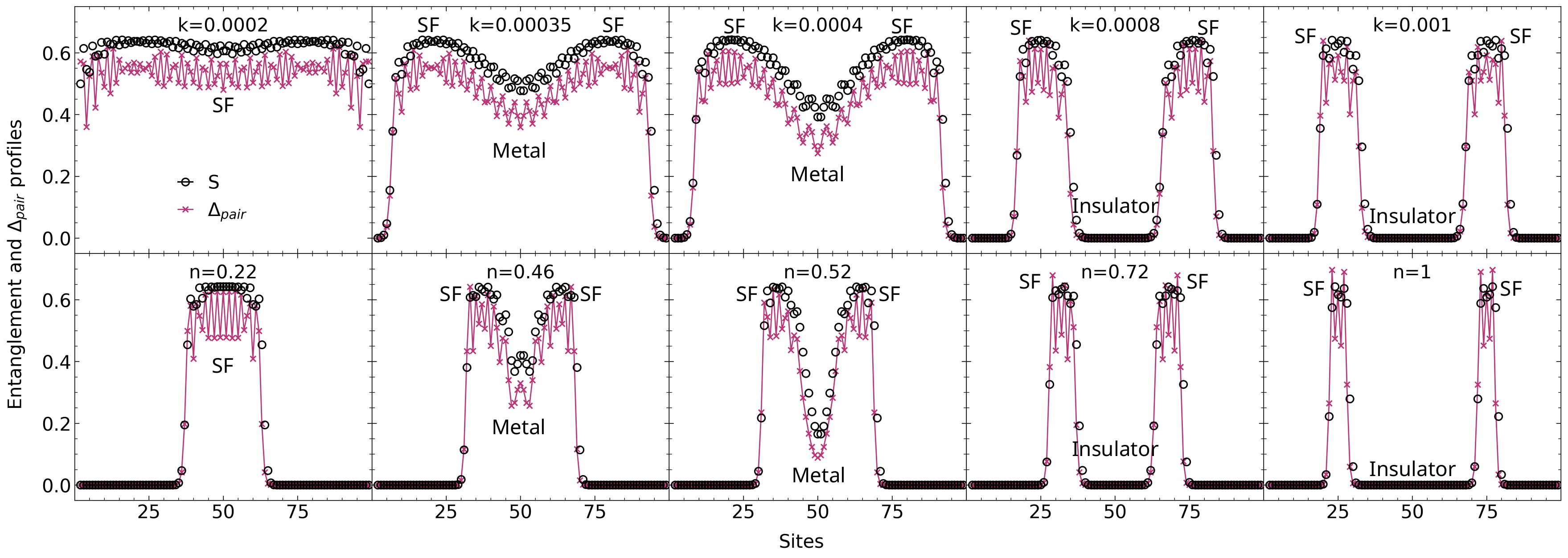}
%\par\end{centering}
\caption{Entanglement $\{S_i\}$ and superconducting order parameter $\{\Delta_{pair}\}$ profiles for several confinement strengths $k$ at fixed $n=1$ (upper panels) and for several average densities $n$ at fixed $k=0.002$ (bottom panels). The initial superfluid (SF), with high entanglement $S_i\sim 0.62$ and $\Delta_{pair}\sim 0.55$ (small $k$ or $n$), evolves to a metallic phase at the bulk (for increasing $k$ or $n$), with finite but reduced $S_i$ and $\Delta_{pair}$, keeping superfluid edges; finally reaching (by increasing further $k$ or $n$) a non-trivial topological insulator (TI), characterized by an insulator bulk surrounded by highly entangled superfluid edges. Notice that both quantities have a semi-quantitative agreement, confirming that entanglement can be used as a topological marker and an order parameter. Here $\Delta_{pair}$ was normalized by a factor $0.73$, such that it resides between $0\leq \Delta_{pair}\leq 1$. In all cases $L=100$ and $U=-10$. \label{fig2}}
\end{figure*}

We start by considering the effects of the confinement potential on the onsite measurements within a superfluid (SF) sample. Figure~\ref{fig1}(a) presents the density profile $\{n_{i}\}$ for several potential curvatures $k$ at a fixed average density $n=1$. We find that the flat charge distribution over the entire chain, observed at the superfluid regime ($k=0$), quickly evolves to a higher concentration of particles at the trap center as the confinement increases. This then defines at the potential core an effective chain \cite{PhysRevB.76.220508, francca2012fulde} {where the fermionic wavefunction extends} (whose size is dependent on $k$), with an effective higher density, $n_{eff}>n$, while the ends of the chain are kept empty. We also see that for $k\gtrsim 0.0005$ there appears a flat plateau with $n_i=2$ (fully occupied sites) at the potential center, which becomes broader as $k$ increases further. This fully occupied bulk characterizes a band-like insulator --- composed of hardcore bosons such as pure Fock states \cite{PhysRevA.70.031603, PhysRevLett.95.220402, PhysRevB.76.220508} --- since within the bulk the charge itinerancy is now totally supressed. Thus the confinement induces a transition at the bulk from a superfluid to an insulator. Notice that the same superfluid-insulator transition can alternatively be induced by the increase of the average density at a fixed $k$, Fig.~\ref{fig1}(b), as also seen in spin-imbalanced chains \cite{PhysRevB.76.220508}, since it leads equivalently to a higher effective density at the bulk.

\begin{figure}[!ht]
%\begin{centering}
\centering
\includegraphics[scale=0.31]{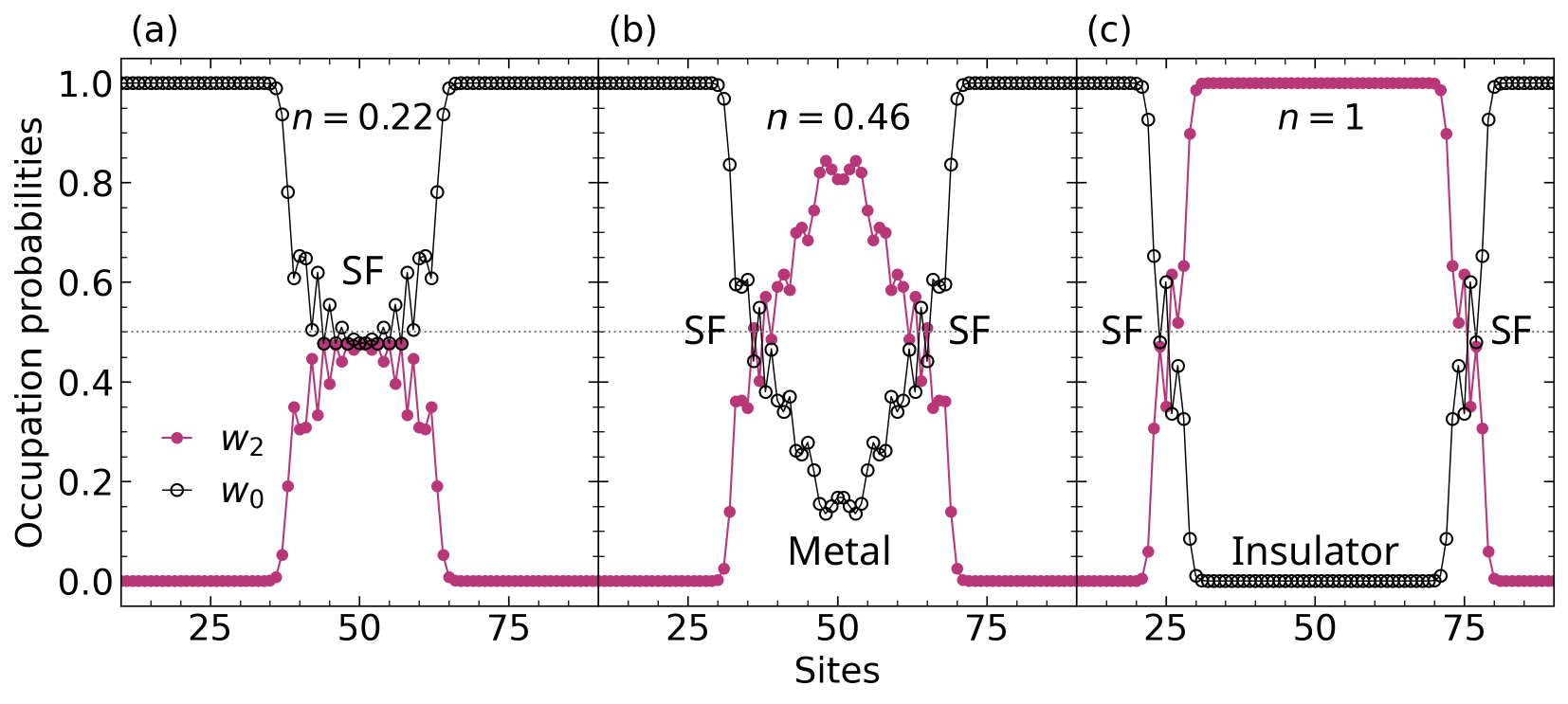}
%\par%\end{centering}
\caption{Double occupancy $\{w_{i2}\}$ and empty-occupation probability $\{w_{i0}\}$ profiles for three distinct average densities within the superfluid (SF), the metallic and the non-trivial topological insulator (TI) regimes. The entanglement at the SF edges is protected by the local particle-hole symmetry ($w_{i2}=w_{i0}$). In all cases $L=100$, $k=0.002$ and $U=-10$.}\label{fig3}
\end{figure}

We then analyze this superfluid-insulator transition through both the single-site entanglement and the superconducting order parameter, as shown in Figure~\ref{fig2}. For small $k$ (or $n$) the bulk is a superfluid, characterized by a high superconducting order parameter ($\Delta_{pair}\approx 0.55$) and high entanglement ($S_i\approx 0.62$). As $k$ (or $n$) increases, the superconducting phase at the bulk is {\it (i)} projected towards the edges ({the outermost sites of the effective region, similarly to the non-interacting case}) and {\it (ii)} replaced by a metallic phase, marked by a smaller but finite entanglement and a corresponding decrease of the $\Delta_{pair}$. By increasing $k$ (or $n$) further, the bulk is transformed into an insulator, vanishing both entanglement and $\Delta_{pair}$. Notice that this intermediate metallic regime could not be distinguished from the density profiles presented in Fig.~\ref{fig1}. It is also interesting that for all cases both quantities, $S_i$ and $\Delta_{pair}$, are semi-quantitatively equivalent{. This is unexpected in principle, since the pair-pair correlation is associated to the creation and annhilation of pairs, thus to $w_2$ and $w_0$; while entanglement is a non-trivial function of all the four probabilities. However, for this strongly attractive regime the single occupation probabilities are negligible, $w_2 + w_0 \approx1$, as shows Figure~\ref{fig3}, thus both $S_i$ and $\Delta_{pair}$ reflect the interplay between double and empty occupancies. T}hus the single-site entanglement can be considered a semi-quantitative order parameter for the insulator, metallic and superfluid phases within these systems.

The most remarkable feature in Fig.~\ref{fig2} is, however, the fact that the {degree of} entanglement is kept essentially constant ($S_i\approx 0.62$) for any $k$ (or $n$): the highly entangled edge states are then robust against perturbations. This robustness of the edge states is a clear signature of symmetry-protected topological phases \cite{PhysRevB.81.064439,RevModPhys.93.045003,PhysRevB.86.125441}. The analysis of the double and empty occupation probabilities, in Figure~\ref{fig3}, reveal that the symmetry protecting entanglement at the edges in this system is the particle-hole symmetry, since locally we have $w_2\sim w_0$. Our findings then show that the superfluid edges are topologically protected by the particle-hole symmetry at the boundaries of the harmonic potential, thus for sufficiently strong confinement the system becomes a {\it non-trivial topological insulator} (TI), characterized by an insulating bulk --- with null $S$ and $\Delta_{pair}$ --- surrounded by highly entangled superfluid edges. The entanglement profile can be used then as a topological marker in theses systems.

\begin{figure}[htbp]
\begin{centering}
\includegraphics[scale=0.56]{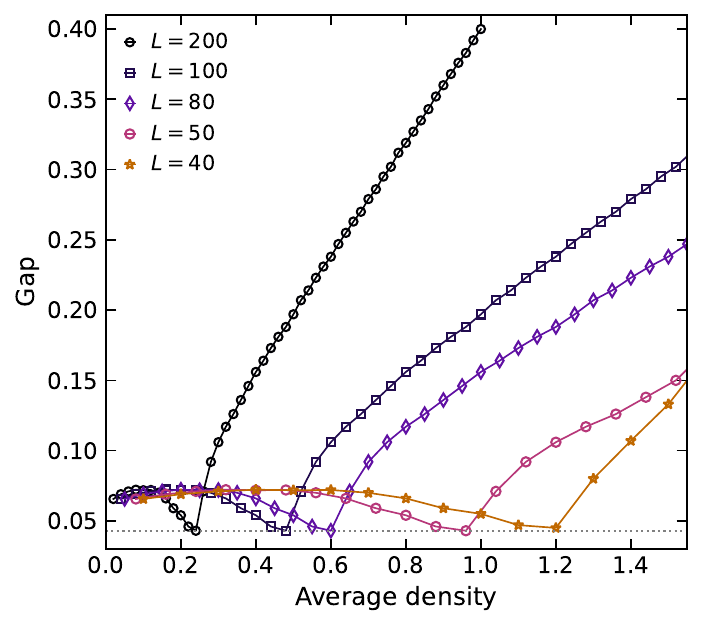}
\par\end{centering}
\caption{Charge gap $\Delta_c$ as a function of the average density $n$ characterizing: {\it i)} the superfluid (SF) phase for $n<n_C$, with a small gap independent on $n$; {\it ii)} the metallic intermediate regime, at $n=n_C$, with $\Delta_c\rightarrow 0$; and {\it iii)} the non-trivial topological insulator (TI) for $n>n_C$, with a larger and increasing with $n$ gap. Notice that increasing $L$ changes $n_C$ and the TI gap, but the SF and the metallic phase gaps are constant and finite in the thermodinamic limit. In all cases $k=0.002$ and $U=-10$. \label{fig4}}
\end{figure}

\begin{figure}[htbp]
\begin{centering}
\includegraphics[scale=1.1]{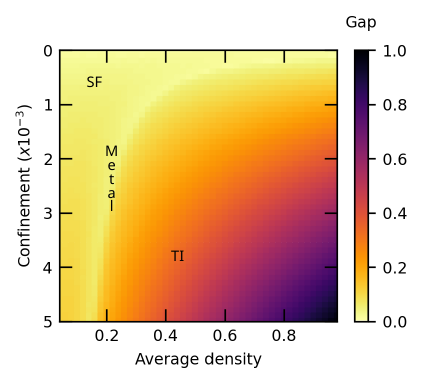}
\par\end{centering}
\caption{Phase diagram (depicted via the charge gap for $L=200$ and $U=-10$) as a function of the confinement strength and average density, demarking the superfluid (SF) phase with finite but small gap, the metallic intermediate regime with $\Delta_c\rightarrow 0$, and the non-trivial topological insulator (TI) phase, with larger gap.\label{fig5}}
\end{figure}

The charge gap $\Delta_c$ in Figure~\ref{fig4} corroborates our interpretation: for a fixed confinement, there exists a critical $n_C$ for which $\Delta_c\rightarrow 0$, corresponding to the metallic bulk regime {(bulk-boundary correspondence)}. For $n<n_C$ the system is a superfluid, with finite but small gap, resembling the pairing gap~\cite{RevModPhys.83.1057}, and almost constant with $n$. While for $n>n_C$ the gap increases almost linearly with $n$, confirming that the system has reached the non-trivial topological insulator (TI) regime. The phase diagram for the SF, metallic and TI phases can be then depicted via the charge gap, as shows Figure~\ref{fig5}. Finally, the analysis of the gap for several chain sizes, Fig.~\ref{fig4}, shows that $\Delta_c$ reaches small but finite values at the metallic state at $n_C$ in the thermodynamic limit. This fact is crucial for the protection of the non-trivial topological properties, and has also been observed in previous investigations of topological Mott insulator in superlattices~\cite{hu2019topological}.

\section{Conclusions}\label{sec_conclusion}
In summary we have investigated the impact of harmonic confinement on the onsite measurements of strongly attractive superfluids. The density profile signs the superfluid to insulator transition at the bulk driven by either the confinement strength (at a fixed average density) or the average density (at a fixed confinement). The profiles of single-site entanglement and superconducting order parameter reveal that the superfluid is projected to the boundaries of the potential, reaching a non-trivial topological insulator, characterized by an insulating bulk surrounded by highly entangled superfluid edges, which are topologically protected by a local particle-hole symmetry. We also find a semi-quantitative agreement between the superconducting order parameter and the single-site entanglement, confirming then that entanglement can be used as a topological marker and an order parameter in these systems. The charge gap not only confirms the transition from a superfluid to a non-trivial topological insulator, but also shows that this transition is mediated by a metallic intermediate regime within the bulk. Using the gap we could depict a phase diagram for which the non-trivial topological insulator can be found in such harmonically confined attractive systems. Our study then shows that confined fermionic chains as described by the one-dimensional Hubbard model can exhibit non-trivial topological insulators with superconducting edges. In the future, if one successfully maps the model to qubits~, it could serve as a platform for quantum information processing.

\begin{acknowledgments}
We thank fruitful discussions with F. Iemini, F. Assaad, G. Diniz, E. Vernek, R. Resta and M. Continentino.  This study was financed by the São Paulo Research Foundation (FAPESP), Brasil (2021/06744-8; 2023/00510-0; 2023/02293-7) and by CNPq (403890/2021-7; 140854/2021-5; 306301/2022-9).
\end{acknowledgments}
\vspace{0.1cm}
{\it Data Availability Statement ---} The data that support the findings of this study are available from the corresponding author upon reasonable request.

%\nocite{*}
%%\bibliography{reference}
\bibliographystyle{apsrev4-2}
\input{revised_2.bbl}

\end{document}

%% file: revised_2.bbl
%aipnum4-2.bst 2019-01-14 (MD) hand-edited version of apsrev4-1.bst